\documentclass[amsmath, amssymb, superscriptaddress, twocolumn, prl]{revtex4-1}

\usepackage{graphicx}
\usepackage{color}
\usepackage{hyperref}

\begin{document}
	
\title{Non-Hermitian topological end breathers}

\author{Li-Jun Lang}
\email{ljlang@scnu.edu.cn}
\affiliation{Guangdong Provincial Key Laboratory of Quantum Engineering and Quantum Materials, School of Physics and Telecommunication Engineering, South China Normal University, Guangzhou 510006, China}
\author{Shi-Liang Zhu}
\affiliation{Guangdong Provincial Key Laboratory of Quantum Engineering and Quantum Materials, School of Physics and Telecommunication Engineering, South China Normal University, Guangzhou 510006, China}
\affiliation{Guangdong-Hong Kong Joint Laboratory of Quantum Matter, Frontier Research Institute for Physics, South China Normal University, Guangzhou 510006, China}
\author{Y. D. Chong}
\email{yidong@ntu.edu.sg}
\affiliation{Division of Physics and Applied Physics, School of Physical and Mathematical Sciences, Nanyang Technological University, Singapore 637371, Singapore}

\date{\today}

\begin{abstract}
  Nonlinearities in lattices with topologically nontrivial band structures can give rise to topological solitons, whose properties differ from both conventional lattice solitons and linear topological boundary states.  We show that a Su-Schrieffer-Heeger-type lattice with both nonlinearity and nonreciprocal non-Hermiticity hosts a novel oscillatory soliton, which we call a topological end breather.  The end breather is strongly localized to a self-induced topological domain near the end of the lattice, in sharp contrast to the extended topological solitons previously found in one-dimensional lattices.  Its stable oscillatory dynamics can be interpreted as a Rabi oscillation between two self-induced topological boundary states, emerging from a combination of chiral lattice symmetry and the non-Hermitian skin effect.  This demonstrates that non-Hermitian effects can give rise to a wider variety of topological solitons than was previously known to exist.
\end{abstract}

\maketitle

Topological solitons are a new and intriguing class of solitons arising from the interplay of nonlinearity with topologically nontrivial band structures \cite{170802-1, 171020-1, 170210-1, 170622-1, 171023-1}.  Whereas conventional solitons can be interpreted as localized states bound to self-induced potential wells \cite{170526-1}, topological solitons utilize a completely different source of localized states, namely the robust topological states occurring at the boundaries of topological materials \cite{111216-2, 120903-2}.  This raises interesting theoretical issues since the concept of band topology is based on linear systems, and the proper way to extend it into the nonlinear regime remains unsettled.  Topological solitons have been found to possess a diverse range of properties that differ significantly from both conventional non-topological lattice solitons \cite{170614-1} and linear topological boundary states.  For instance, in a Su-Schrieffer-Heeger (SSH) model \cite{130704-1}---a prototypical one-dimensional (1D) lattice model with two topologically distinct phases---introducing nonlinearity can generate an end soliton that is evidently related to the linear SSH model's topological end state, but behaves in strikingly different ways \cite{170622-1, 171023-1, 180316-1}; in particular, the wave function forms a ``plateau'' extending into the bulk rather than decaying exponentially to zero \cite{170622-1}.  Two-dimensional (2D) lattices, on the other hand, support topological solitons that are exponentially localized, either circulating around a point in the bulk~\cite{170802-1} or traveling unidirectionally along the lattice edge \cite{171020-1, 170210-1}, and that can exhibit nontrivial pseudospin textures \cite{Marzuola2019}.  The first experimental realizations of topological solitons have only recently been achieved, using nonlinear electrical circuit lattices \cite{180316-1, 190307-1} and nonlinear waveguide arrays \cite{Maczewsky2020, Mukherjee2020, Xia2020, Zhang2020}.   These novel solitons may have interesting roles to play in applications such as topological lasing \cite{180402-1, 180409-1, 180719-1, 180720-9, 180720-10, Wang2020}, topologically enhanced harmonic generation \cite{190307-1,Kruk2019}, and topological synchronization \cite{Kotwal2019, Sone2020}.

Another interesting way to modify soliton properties is through non-Hermitian effects \cite{110707-1, 200911-1}, such as gain and loss \cite{Chen2020, Bloch2021}, which can give rise to a variety of phenomena such as the spontaneous breaking of gain/loss symmetries \cite{110707-1}.  The consequences of non-Hermiticity for the topological properties of linear lattices is an active area of research \cite{200911-1}.  One of the most interesting phenomena that has been discovered is the non-Hermitian skin effect~\cite{181022-1, 181221-1, Okuma2020}, a breakdown of the bulk-boundary correspondence principle that induces bulk states to collapse to the boundaries of a non-Hermitian lattice \cite{171012-1, 181022-1, 180720-4, 180720-6, 181016-1, Torres18, 180720-7, Rudner2009, 181221-1, Esaki2011, 180720-8, Lieu18, 180720-3, 181019-1, 181221-2, 181127-1, Lee2019, 180924-1, 190106-1, Wang2019a, Borgnia2020,  Okuma2020, Fang2020}.  This has been experimentally observed in diverse platforms including photonics \cite{Xue2020, Weidemann2020, Fan2020}, cold atoms~\cite{Yan2020}, electric circuits \cite{Thomal2020, Hofmann2020}, and mechanical lattices \cite{Brandenbourger2019, Ghatak2020}.

Here, we show that a non-Hermitian 1D lattice with saturable nonlinearity can support a self-sustained oscillatory mode that we call a ``topological end breather.''  Its properties are distinct from the end soliton found in the Hermitian nonlinear SSH model \cite{170622-1, 171023-1, 180316-1}.  The end soliton of the Hermitian model is a stationary mode forming a long-tailed plateau that ``smears'' the topologically nontrivial domain across the entire lattice.  The topological end breather, by contrast, is a localized oscillation within a self-induced topological domain that forms near the lattice edge, with a different topological phase from the rest of the lattice.  The formation of a domain wall is aided by the non-Hermitian skin effect \cite{181022-1, 181221-1, Okuma2020, 171012-1, 181022-1, 180720-4, 180720-6, 181016-1}, which suppresses intensities in the topologically trivial bulk.  The topological end breather is thus even more strongly localized than the topological end states of the linear Hermitian SSH model, similar to the pointlike topological solitons found in 2D Hermitian lattices \cite{170802-1, 171020-1, 170210-1}, and in contrast to the extended solitons previously seen in 1D \cite{170622-1}.

Unlike the discrete breathers that have long been known to exist in nonlinear lattices \cite{170704-1},
the topological end breather derives its behavior from the topological properties of the underlying lattice.  The conditions for its emergence are given by the saturated and unsaturated values of a nonlinear parameter relative to a critical value set by the non-Hermitian topological transition.  We show that the stable oscillatory behavior can be interpreted as a Rabi oscillation between two states in the topologically nontrivial domain, and that the spatial width of this domain determines the oscillation period.  Unlike other recently studied nonlinear non-Hermitian lattice models \cite{Kotwal2019, Sone2020}, there is negligible bulk dynamics due to the non-Hermitian skin effect.  We will also discuss possible experimental approaches for realizing topological end breathers \cite{Xue2020, Weidemann2020, Fan2020, Yan2020, Thomal2020, Hofmann2020, Brandenbourger2019, Ghatak2020}.

\begin{figure}
  \includegraphics[width=1\linewidth]{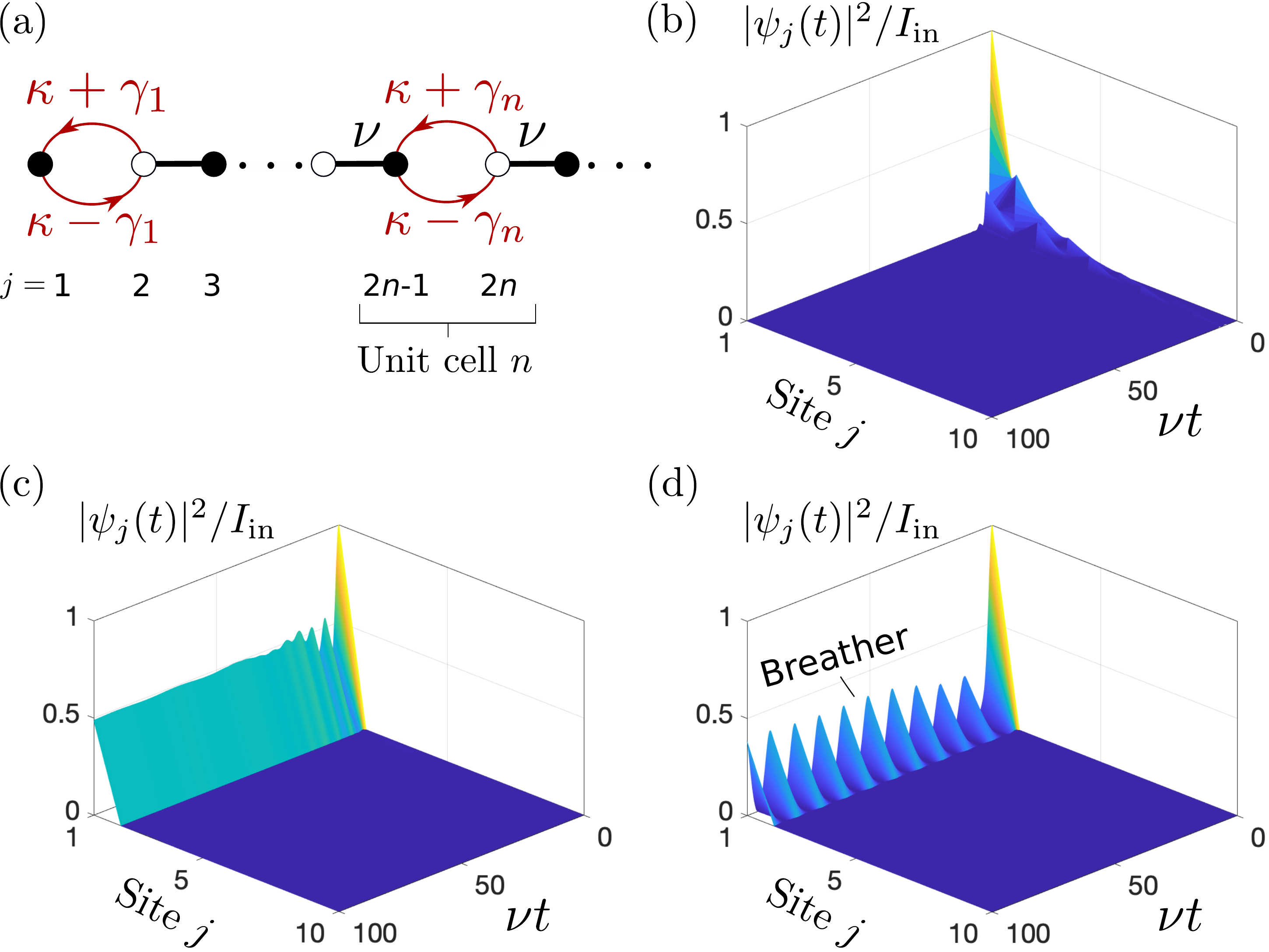}
  \caption{(a) Schematic of nonlinear nonreciprocal SSH model with OBC. In the $n$th cell, the nonlinear hopping $\gamma_n$ depends on the intensity at sites $2n-1$ and $2n$.  (b)--(d) Time evolution of site intensities for an initial state $\psi(0)=(\sqrt{I_{\rm in}},0,0,\cdots)$, with $I_{\textrm{in}} = 10^3 I_s$, $\kappa = \sqrt{2}\nu$, and different values of $\gamma_0$ and $\gamma_s$.  The calculations are done with $100$ cells, but only the first ten sites are plotted.  (b) For $\gamma_0 = 0$ and $\gamma_s = 0.5\nu$, the excitation decays rapidly to zero.  (c) For $\gamma_0 = 1.2\nu$ and $\gamma_s = (\sqrt{7}/2) \nu$, a steady-state end mode is excited.  (d) For $\gamma_0 = 0$ and $\gamma_s = (\sqrt{7}/2)\nu$, a topological end breather appears at the end of the lattice. }
\label{fig1}
\end{figure}

We consider a nonlinear nonreciprocal SSH-like model, shown schematically in Fig.~\ref{fig1}(a), with the Hamiltonian
\begin{align}
  H = \sum_{n = 1, 2, \dots} &\Big[(\kappa+\gamma_n)a^\dag_{2n-1}
    a_{2n} + (\kappa-\gamma_n)a^\dag_{2n}a_{2n-1} \notag \\
    &+ \nu (a^\dag_{2n}a_{2n+1}+a^\dag_{2n+1}a_{2n})\Big], 
\label{nln_Ham}
\end{align}
where ${a}_j^{(\dag)}$ is the annihilation (creation) operator on site $j$, $n$ is the unit cell index, and $\kappa$ and $\nu$ are real hopping parameters.  The intercell hoppings $\nu$ are reciprocal; the intracell hoppings are in general nonreciprocal, with hopping strengths $\kappa\pm\gamma_n$ in opposite directions.  Each $\gamma_n$ is a nonlinear hopping coefficient described below.

If each $\gamma_n$ is set to a real constant $\gamma$, the model reduces to the nonreciprocal SSH model \cite{181022-1}.  Under open boundary conditions (OBCs), this non-Hermitian model exhibits end states that can be continued to the topological end states of the Hermitian ($\gamma=0$) SSH model.  One finds that $|\kappa^2-\gamma^2| < \nu^2$ corresponds to a nontrivial phase hosting topological end states \cite{181022-1}, and that the non-Hermitian skin effect localizes all bulk states to one end of an open finite chain---a phenomenon that can be interpreted as a breakdown of Bloch's theorem.  The localized states are topological in origin, in that they can be mapped to topological end states of a family of auxiliary Hermitian Hamiltonians \cite{Okuma2020}.

Returning to the nonlinear case, we let the nonlinear hopping terms have the saturable form
\begin{align}
  \gamma_n &= \gamma_s - \frac{\gamma_s - \gamma_0}{1+I_n(t)/I_s},
  \label{e-sat_nln} \\
  I_n(t) &\equiv |\psi_{2n-1}(t)|^2+|\psi_{2n}(t)|^2,
\end{align}
where $I_s$ is a saturation intensity scale, $\gamma_0$ and $\gamma_s$ are real constants, and $\psi_j(t)$ denotes the single-particle wavefunction on site $j$.  Thus, $\gamma_n \rightarrow \gamma_0$ in the linear (zero-intensity) limit and $\gamma_n \rightarrow \gamma_s$ in the high-intensity limit.

Although nonlinear hoppings may be difficult to realize in physical systems, this model can be mapped by a unitary transformation to a Cruetz model with saturable on-site gain/loss, which may be easier to realize \cite{180919-1P}.  For details, see the Supplemental Material \cite{sm}.

\begin{figure*}
  \includegraphics[width=\textwidth]{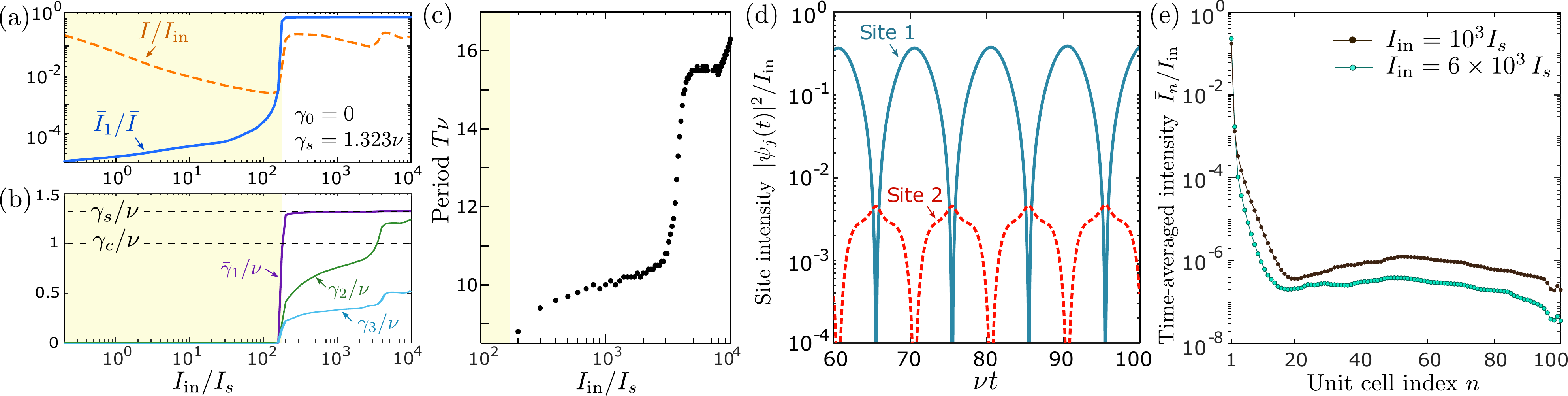}
  \caption{(a) Time-averaged intensities vs normalized input intensity $I_{\mathrm{in}}/I_s$.  The lattice has 100 cells with $\gamma_0 = 0$ and $\gamma_s = (\sqrt{7}/2)\nu$, as in Fig.~\ref{fig1}(d).  The blue solid curve shows $\bar{I}_1 / \bar{I}$ and the orange dashed curve shows $\bar{I} / I_{\mathrm{in}}$, where $\bar{I}_n$ is the time-averaged intensity in cell $n$ and $\bar{I} = \sum_n \bar{I}_n$.  The time averages are calculated over $50/\nu \le t \le 100/\nu$.  With increasing input intensity, $\bar{I}_1 / \bar{I}$ goes from nearly zero to unity due to the emergence of the end breather.  (b) Plots of $\bar{\gamma}_{1,2,3}/\nu$ vs $I_{\mathrm{in}}/I_s$, revealing that the appearance of the end breather in (a) corresponds to the transition from $\bar{\gamma}_1 < \gamma_c$ (yellow background) to $\bar{\gamma}_1 > \gamma_c$ (white background).  A subsequent uptick in $\bar{I}$ corresponds to $\bar{\gamma}_2$ crossing the critical point.  (c) Period of the end breather vs $I_{\mathrm{in}}/I_s$.  (d) Evolution of the site intensities in the first cell for $I_{\mathrm{in}} = 10^3 I_s$.  (e) Time-averaged intensity $\bar{I}_n/I_{\mathrm{in}}$ vs cell index $n$, calculated at $t = 100/\nu$ for $I_{\mathrm{in}} = 10^3 I_s$ (black circles) and $I_{\mathrm{in}} = 6\times10^3 I_s$ (cyan circles). }
\label{fig2}
\end{figure*}

In the following, we take $\kappa > \nu>0$, so that the lattice is topologically trivial in the linear limit, and $0 \le \gamma_0 \le \gamma_s \le \kappa$, so that the nonlinear intracell hoppings are non-negative.  In the linear regime, the non-Hermitian skin effect concentrates eigenstates of the finite chain (with OBC) on the left boundary.  In the nonlinear case, the behavior depends on the choice of $\gamma_0$ and $\gamma_s$ relative to the critical value $\gamma_c = \sqrt{\kappa^2 - \nu^2}$.  Let the leftmost site be initially excited: $\psi_j(0) = \sqrt{I_{\textrm{in}}} \, \delta_{j1}$, where $j = 1, \dots, 2N$.  The state then evolves according to the nonlinear Hamiltonian \eqref{nln_Ham}.  Note that $\sum_n I_n(t)$ is not conserved in the non-Hermitian regime, but the initial excitation generally does not lead to an exponential blow-up at long times since the instantaneous Hamiltonian of the finite chain is pseudo-Hermitian \cite{200911-1,171012-1,181022-1}.  

The dynamical behavior in the linear limit has previously been studied \cite{210108-1}.  For $\gamma_0 < \gamma_c$, the initial excitation at $j = 1$ decays away in time, consistent with the topological triviality of the corresponding lattice.  For $\gamma_0 > \gamma_c$, the excitation creates a long-lived mode on the left boundary, which can be attributed to the existence of the topological end state and the suppression of diffraction in the bulk by the non-Hermitian skin effect \cite{210108-1}.

For the nonlinear case, there are three regimes of interest, depending on the choice of $\gamma_0$ and $\gamma_s$.  First, for $\gamma_0 \le \gamma_s < \gamma_c$, all values of $\gamma_n$ lie below the critical value, and the initial excitation decays away, similar to the linear topologically trivial case, as shown in Fig.~\ref{fig1}(b).  Second, for $\gamma_c < \gamma_0 \le \gamma_s$, the initial excitation forms a steady-state solution localized on the boundary, similar to the linear topological case, as shown in Fig.~\ref{fig1}(c).

Most interesting of all is the third case, $\gamma_0 < \gamma_c < \gamma_s$, which is the focus of this paper.  As shown in Fig.~\ref{fig1}(d), the excitation produces a self-sustained oscillation near the lattice boundary, which we call an ``end breather.''  Conceptually, regions of the lattice with low intensity are topologically trivial, and regions with high intensity are nontrivial; the competition between the two phases allows for a self-induced boundary state with nontrivial dynamics.  We shall see that the non-Hermitian skin effect plays a crucial role in this, and that the resulting breather mode is distinct from the previously studied end solitons of Hermitian nonlinear SSH models \cite{170622-1}.

To demonstrate that the end breather is a self-induced topological mode, Fig.~\ref{fig2}(a) plots the time-averaged relative intensities versus input intensity.  The blue solid curve shows $\bar{I}_1 / \bar{I}$, where $\bar{I}_1$ is the time-averaged intensity in cell $n = 1$ and $\bar{I} = \sum_n \bar{I}_n$ is the time-averaged intensity summed over all cells; the orange dashes show $\bar{I} / I_{\mathrm{in}}$.  The time averages are taken over an interval much longer than the end breather's period, after transients have died away.  As $I_{\textrm{in}}$ increases past a threshold value, $\bar{I}_1 / \bar{I}$ increases abruptly from very low levels to a saturated value around unity, signifying that the intensities become concentrated in the leftmost cell due to the formation of the end breather.  In Fig.~\ref{fig2}(b), we see that this occurs precisely when $\bar{\gamma}_1 > \gamma_c$, where $\bar{\gamma}_1$ is the time-averaged nonlinear hopping term in the first cell.

Figure~\ref{fig2}(c) shows the variation of the end breather's period with $I_{\mathrm{in}}$.  Above threshold, the period increases gradually with $I_{\mathrm{in}}$ until there is an abrupt increase at $I_{\mathrm{in}} \approx 4000\, I_s$.  The abrupt change occurs when $\bar{\gamma}_2$ exceeds $\gamma_c$ [see Fig.~\ref{fig2}(b)], and is associated with a shift in the end breather's self-induced topological domain wall.  In Fig.~\ref{fig2}(d), we plot the periodic time dependence of the site intensities in the first cell for the exemplary case of $I_{\mathrm{in}} = 10^3 I_s$.  The time-averaged intensities after time $t = 100/\nu$ are plotted in Fig.~\ref{fig2}(e), revealing strong localization to the lattice boundary.

\begin{figure}
  \includegraphics[width=1\linewidth]{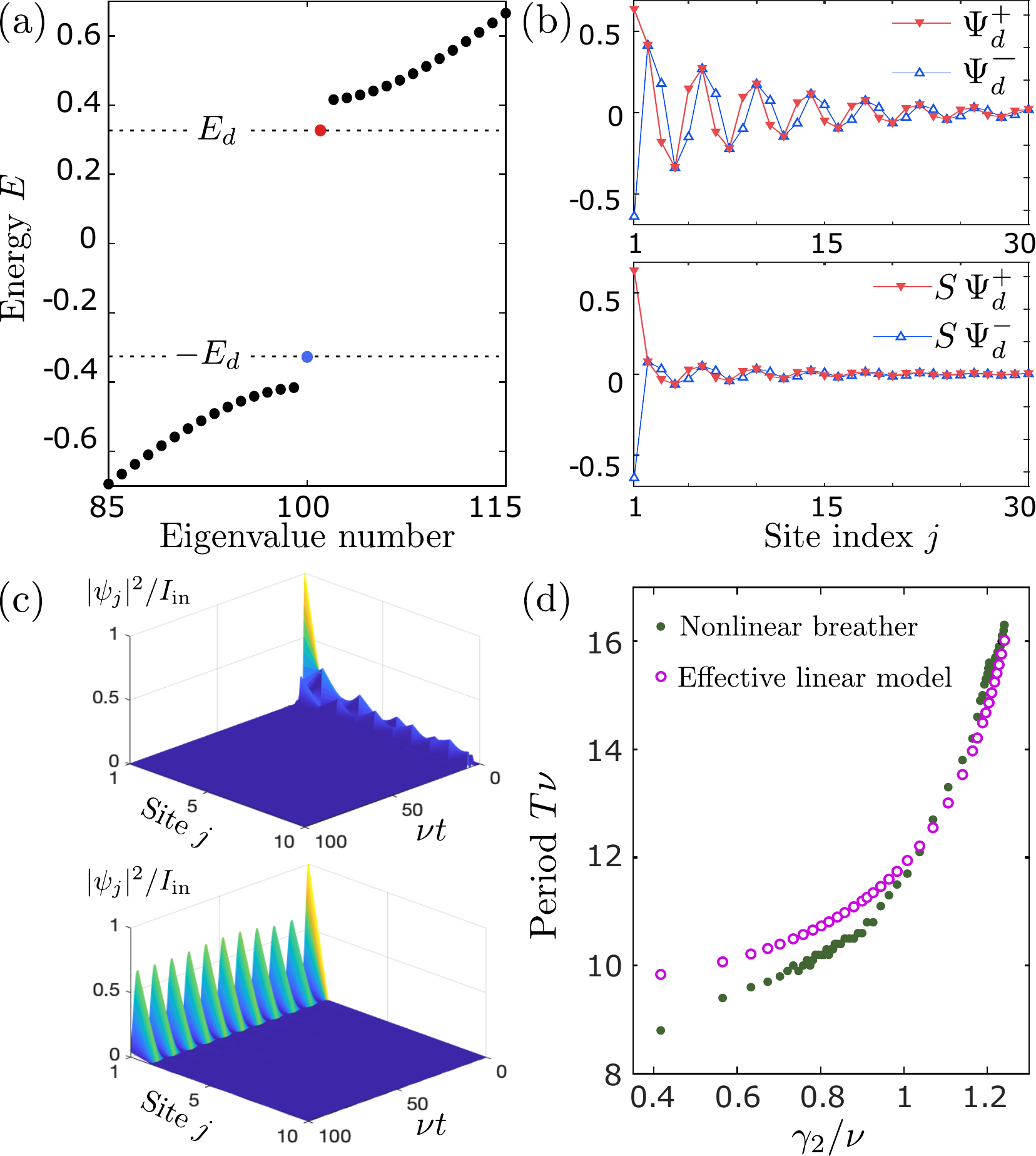}
  \caption{(a) Spectrum of the non-Hermitian Hamiltonian $H_l$ with parameters $\kappa=\sqrt{2}\nu$, $\gamma_1 =\gamma_s= (\sqrt{7}/2)\nu$, $\gamma_n=\gamma_0=0$ for $n > 1$, and $100$ cells.  The chirality-paired defect states, with energies $\pm E_d$, are plotted in red and blue.  (b) Upper panel: Spatial distribution of the defect states $\Psi_d^+$ (red) and $\Psi_d^-$ (blue) vs site index $j$.  Lower panel: Spatial distribution of $S\Psi_d^+$ (red) and $S\Psi_d^-$ (blue).  (c)  Time evolution of site intensities in the linear model, given the same initial state as in Fig.~\ref{fig1}, with the same settings as (a), but with (upper panel) $\gamma_1 =\gamma_d= 0.5\nu$, and (lower panel) $\gamma_1 =\gamma_s= (\sqrt{7}/2)\nu$.  (d) Period of the nonlinear breather, plotted against the time-averaged value of $\gamma_2 / \nu$ (gray dots); and the period predicted by the effective linear model using $\gamma_1 = \gamma_s$, varying $\gamma_2$, and $\gamma_{n > 2} = 0$ (purple circles).  The data for the end breather are extracted from time-domain simulations using input intensities $I_{\mathrm{in}} \in [200, 10^4] I_s$, similar to Fig.~\ref{fig2}(c).}
\label{fig3}
\end{figure}

Although the intensities oscillate between different sites, as seen in Fig.~\ref{fig2}(c), the domain wall is stationary over each period; see Fig.~S2(d) in the Supplemental Material \cite{sm}.  We can hence model its effects with a static linear non-Hermitian chain, similar to Fig.~\ref{fig1}(a) but with intensity-independent $\gamma_n$'s.  The linear model's Hamiltonian $H_l$ can be related to a Hermitian linear Hamiltonian $H_h$ by $H_l = S H_h S^{-1}$, where the similarity matrix $S$ is given in the position basis by
\begin{align}
  \begin{aligned}
    S &= \mathrm{diag}[1, \beta_1, \beta_1, \beta_1\beta_2, \beta_1\beta_2, \dots], \\
    \beta_n &= \sqrt{\frac{\kappa-\gamma_n}{\kappa+\gamma_n}}.
  \end{aligned}
  \label{Sop}
\end{align}
Note that $S$ is not unitary.  Each eigenstate $|\psi\rangle$ of $H_h$ maps to an eigenstate $S|\psi\rangle$ of $H_l$ with the same energy.  Since $\beta_n \le 1$, applying $S$ to $|\psi\rangle$ results in a relative decrease in the site intensities away from the left boundary, giving rise to a non-Hermitian skin effect for $H_l$ \cite{181022-1, 210108-1}.

To model the breather just above its threshold, we set $\gamma_1\equiv\gamma_d$ (beyond the threshold, $\gamma_d$ quickly saturates to $\gamma_s$) and $\gamma_n = \gamma_0$ for all $n \ge 2$.  The corresponding $H_h$ describes an SSH chain containing an end defect generated by an intracell hopping $\kappa_d = \sqrt{\kappa^2-\gamma_d^2}$ in the $n=1$ cell, with intracell hopping $\kappa_0 = \sqrt{\kappa^2-\gamma_0^2}$ in all other cells.  When $\gamma_d > \gamma_c$, we have $\kappa_d < \nu < \kappa_0$, so the defect can be regarded as a one-cell-wide domain with a different topological phase from the rest of the chain.  The defect induces a pair of end states $|\Psi_d^\pm\rangle$ with energies $\pm E_d$, as shown in Fig.~\ref{fig3}(a).  These originate from the two topological end states of the SSH domain formed by the defect.  As the domain is just one cell wide, the end state energies undergo strong level repulsion away from zero.  Since $\pm E_d$ lie in the gap, the wave functions decay exponentially into the bulk, as shown in the upper panel of Fig.~\ref{fig3}(b).  When $\gamma_d<\gamma_c$, the end states disappear since the one-cell defect lies in the topologically trivial regime, as the rest of the lattice. Further details of the defect states are given in the Supplemental Material \cite{sm}.

Note that the linear Hamiltonians $H_l$ and $H_h$, as well as the nonlinear Hamiltonian $H$, all anticommute with the chiral operator $C = \mathrm{diag}(-1,1,-1,1,\cdots)$.  The chiral symmetry ensures that the defect states are related by $|\Psi_d^+\rangle = C |\Psi_d^-\rangle$, with energies $\pm E_d$.

Starting from $|\psi(0)\rangle$, the subsequent state of the non-Hermitian chain is $|\psi(t)\rangle = Se^{-iH_{\rm h}t}S^{-1} |\psi(0)\rangle$.  As $|\psi(0)\rangle$ is localized entirely to $j=1$, it is unaffected by $S^{-1}$.  The initial excitation overlaps strongly with the defect states $|\Psi^\pm_d\rangle$ (see Supplemental Material \cite{sm}).  Hence,
\begin{equation}
  |\psi(t)\rangle \approx \sum_\pm  c_\pm e^{\mp iE_dt} S|\Psi_d^\pm\rangle
  \label{e-approx_evol}
\end{equation}
for some coefficients $c_\pm$.  This describes a Rabi oscillation with period $T_d = \pi/E_d$.  The solution is strongly localized to the boundary due to the action of $S$ (i.e., the non-Hermitian skin effect), as shown in the lower panel of Fig.~\ref{fig3}(b).  The oscillatory behavior of the linear effective model, shown in the lower panel of Fig.~\ref{fig3}(c), is very similar to the end breather [Fig.~\ref{fig1}(d)].  For $\gamma_d < \gamma_c$, however, the end states do not exist in the linear model and no oscillation is observed, as shown in Fig.~\ref{fig3}(c), similar to the nonlinear case [Fig.~\ref{fig1}(b)].

The above analysis applies close to the threshold of the end breather, but it can be generalized to the above-threshold regime, where the first-cell hopping is nearly saturated ($\gamma_1=\gamma_s$).  With increasing intensity, the time-averaged $\gamma_2$ becomes non-negligible, and eventually saturates [Fig.~\ref{fig2}(b)].  When $\gamma_2$ exceeds $\gamma_c$, the domain formed by the defect becomes two cells wide.  In the effective linear model, Eq.~\eqref{Sop} remains valid and $H_l$ and $H_h$ retain their chiral symmetry; with increasing $\gamma_2$ (keeping $\gamma_1 = \gamma_s$ and $\gamma_{n>2} = 0$), the effective linear model exhibits a decrease in $E_d$ (due to the widening of the domain, which reduces level repulsion), so $T_d$ increases.  As shown in Fig.~\ref{fig3}(d), the values of $T_d$ produced by the effective linear model agree well with the end breather periods extracted from time-domain simulations.  The spatial distribution also becomes more localized to the defect with increasing intensity, due to the reduced level repulsion in the wider defect domain and the stronger skin effect due to smaller $\beta_2$.  This is consistent with the behavior of the nonlinear model shown in Fig.~\ref{fig2}(e).

In conclusion, a strongly localized topological end breather can exist in a non-Hermitian nonlinear 1D lattice, arising from a self-induced topological domain.  This phenomenon is tied to the non-Hermitian skin effect \cite{sm}, and does not occur in other nonlinear SSH-like chains (for instance, applying a Hermitian nonlinearity to the hoppings yields an end soliton with a plateaulike form, not a self-induced domain wall \cite{170622-1}).  As the non-Hermitian skin effect is directional, the end breather only exists at one of the two boundaries of the chain; moreover, it cannot occur in the bulk unless there is a preexisting domain wall in the distribution of the nonreciprocal couplings.  To realize these breather modes, it may be helpful to use an equivalent Creutz model \cite{131104-2, 171012-1} with reciprocal linear hoppings and saturable on-site gain/loss, as discussed in the Supplemental Material \cite{sm}; such a lattice could be implemented using cold atoms \cite{Yan2020,180919-1P,Zhang2018}, photonics \cite{Xue2020,Weidemann2020,Fan2020}, electric circuits \cite{180316-1,190307-1,Thomal2020,Hofmann2020}, or mechanical lattices~\cite{Brandenbourger2019,Ghatak2020}.  In the future, it would be interesting to find topological solitons induced by different combinations of non-Hermiticity with nonlinearity.  The non-steady-state behavior of the topological end breather may also have interesting connections to dynamical phenomena such as quenching and dynamical phase transitions \cite{120323-1, Heyl2018, LeeSong2020}.

\begin{acknowledgments}
L.-J.L.~was supported by the National Natural Science
Foundation of China (Grant No.~11904109), the Guangdong Basic and Applied Basic Research Foundation (Grant No.~2019A1515111101), and the Science and Technology Program of Guangzhou (Grant No.~2019050001);
S.-L.Z.~was supported by the Key-Area Research and Development Program of Guangdong Province (Grant No. 2019B030330001) and the National Natural Science Foundation of China (Grants No. 12074180 and No. U1801661);
C.Y.D.~was supported by Singapore Ministry of Education Academic Research Fund Tier 3 Grant No. MOE2016-T3-1-006, Tier 1 Grants No. RG187/18(S) and No. RG148/20, and Tier 2 Grant No. MOE2019-T2-2-085.
\end{acknowledgments}

\bibliography{ref}
\bibliographystyle{apsrev4-1}

\clearpage
\begin{widetext}

\makeatletter 
\renewcommand{\theequation}{S\arabic{equation}}
\makeatother
\setcounter{equation}{0}

\makeatletter 
\renewcommand{\thefigure}{S\@arabic\c@figure}
\makeatother
\setcounter{figure}{0}

\makeatletter 
\renewcommand{\thesection}{S\arabic{section}}
\makeatother
\setcounter{section}{0}

\begin{center}
  {\Large \textbf{Supplemental Material}}
\end{center}

\section{Analytical solution for end states in an effective linear model}

In this section, we analytically solve the end states of the linear model described in the main text.  The intra-cell hoppings are taken to be $\gamma_1=\gamma_d$ and $\gamma_{n}=\gamma_0$ for $n\ge 2$.  For simplicity, we normalize the inter-cell hopping to $\nu=1$.  The non-Hermitian linear Hamiltonian $H_l$ is related to a Hermitian linear Hamiltonian $H_h$ by $H_l=SH_hS^{-1}$, where $S$ is defined in Eq.~(4) of the main text.

Consider end states for $H_h$ of the form
\begin{equation}
  \left|\Psi_d\right\rangle = \mathcal{N}^{-1}\; \big(1,b,a, b r^{-1},ar^{-2},\cdots\big)^T,
\end{equation}
where $\mathcal{N}$ is a normalization factor.  From the static Schr\"{o}dinger equation $H_h|\Psi_d\rangle=E|\Psi_d\rangle$, we obtain
\begin{align}
  \begin{aligned}
  E &= \kappa_d b, &
  Eb &=\kappa_d +a, \\
  Ea &=\kappa_0 br^{-1}-b, &
  Eb &=\kappa_0 a r-a,
  \end{aligned}
\end{align}
where $\kappa_d\equiv\sqrt{\kappa^2-\gamma_d^2}$ and $\kappa_0\equiv\sqrt{\kappa^2-\gamma_0^2}$.  Hence, we can rewrite the end states as
\begin{equation}
  \left|\Psi_d^\pm\right\rangle
  = \mathcal{N}^{-1} \big(\pm 1,b,\pm a, b r^{-1},\pm ar^{-2},\cdots\big)^T,
  \label{end_state}
\end{equation}
where
\begin{equation}
  a =\frac{\kappa_d }{\kappa_0^2-\kappa_d}, \quad
  b =\sqrt{1-\frac{1}{\kappa_0^2-\kappa_d^2}} \quad
  r =\kappa_0 - \frac{\kappa_d^2}{\kappa_0}.
  \label{parameter}
\end{equation}
The normalization factor can be shown to satisfy
\begin{equation}
  \mathcal{N}^2 = 1+(b^2+a^2)(1+r^{-2}+r^{-4}+\cdots)
  =1+\frac{b^2+a^2}{1-r^{-2}}
  = \frac{2(\gamma_d^2-\gamma_0^2)(1-\gamma_d^2+\gamma_0^2)}
  {\kappa^2-\gamma_d^4-\gamma_0^4-\gamma_0^2(1-2\gamma_d^2)}.
\end{equation}
The corresponding eigenenergies are
\begin{equation}
  E_\pm \equiv \pm E_d
  = \pm \kappa_d\sqrt{1-\frac{1}{\kappa_0^2-\kappa_d^2}}
  = \pm  \sqrt{(\kappa^2-\gamma_d^2)
    \left(1-\frac{1}{\gamma_d^2-\gamma_0^2}\right)}.
\end{equation}

\begin{figure}
  \includegraphics[width=1\linewidth]{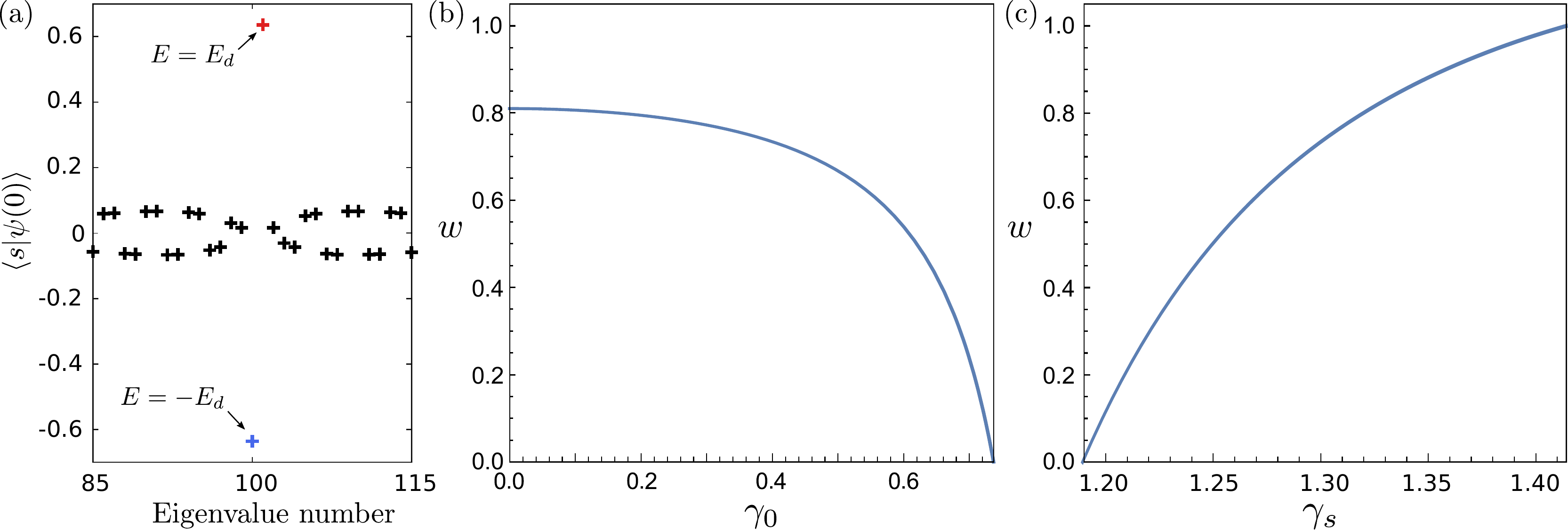}
  \caption{(a) Values of $\langle s | \psi(0)\rangle$, where $|s\rangle$ is the $s$-th eigenstate of the linear effective Hamiltonian and $|\psi(0)\rangle = (1, 0, 0, \dots)$ is the initial state.  Note that the eigenfunctions are all real.  The two eigenstates with the largest overlaps correspond to the defect states with energies $E = \pm E_d$; see also Fig.~3(a) of the main text.  (b) $w$ versus $\gamma_0\in(0,\gamma_0^c)$ for $\kappa=\sqrt{2}$ and $\gamma_s=\sqrt{7}/2$; (c) $w$ versus $\gamma_s\in(\gamma_s^c,\kappa)$ for $\kappa=\sqrt{2}$ and $\gamma_0=0.$}
\label{figs2}
\end{figure}

From Eq.~\eqref{end_state}, we can see that $|\Psi_d^+\rangle = C |\Psi_d^-\rangle$, where $C=\mathrm{diag}(-1,1,-1,1,\cdots)$ is the chiral operator.  We can also check that the parameters $a$, $b$, and $r$ in Eq.~\eqref{parameter} are all real, so the wavefunctions are real.  Moreover, for $r>1$ the wavefunctions are localized to the left boundary---i.e., they are valid end states. 

When we regard $\gamma_d=\gamma_s$ (and thus $\kappa_d=\kappa_s$) as the threshold criterion (due to the fast saturation after $\gamma_d>\gamma_c$), the condition for $r>1$ requires that given $\kappa_0$ or $\gamma_0$, 
\begin{equation}
  \kappa_s\le\sqrt{\kappa_0^2-\kappa_0}\equiv \kappa_s^c \qquad
  \text{or} \quad
  \gamma_s\ge \sqrt{\gamma_0^2+\sqrt{\kappa^2-\gamma_0^2}}\equiv \gamma_s^c,
\end{equation}
and given $\kappa_s$ or $\gamma_s$,
\begin{equation}
  \kappa_0\ge \frac{1}{2}{\kappa_s^2+\frac{1}{4}}\equiv \kappa_0^c \qquad
  \text{or} \quad
  \gamma_0 \le
  \sqrt{\gamma_s^2-\frac{1}{2}-\sqrt{\kappa^2-\gamma_s^2+\frac{1}{4}}}
  \equiv \gamma_0^c.
\end{equation}
Note that $\gamma_s^c\ne \gamma_0^c\ne\gamma_c=\sqrt{\kappa^2-1}$.  We can understand this as follows: if the domain wall were infinitely far away from the end of the lattice, $\gamma_c$ would be the exact critical point.  But since the domain is only one unit cell wide, its two end states hybridize strongly.

Given an initial state, say $|\psi(0)\rangle=(1,0,0,\cdots)$, the overlaps with the two end states are given by
\begin{equation}
  \big\langle\Psi_d^\pm \big| \psi(0) \big\rangle = \pm \, \mathcal{N}^{-1}.
\end{equation}
As shown in Fig.~\ref{figs2}(a), the overlaps with all the other eigenstates are much smaller in magnitude.  This justifies the approximate formula for $|\psi(t)\rangle$ given in Eq.~(5) of the main text, with $c_\pm=\pm\, \mathcal{N}^{-1}$.

Hence, from Eq. \eqref{end_state},
\begin{equation}
  \begin{pmatrix}
    \psi_{2n-1}(t) \\ \psi_{2n}(t)
  \end{pmatrix}
  = \frac{2}{\mathcal{N}} \prod_{m=0}^{n-1}\beta_{m}
  \begin{pmatrix} \cos (E_dt) \\ -i\beta_n\sin (E_dt)
  \end{pmatrix}, \quad n = 1, 2, \dots \;\;\;
  \mathrm{where}\;\;\;
  \beta_n = \sqrt{\frac{\kappa-\gamma_n}{\kappa+\gamma_n}}.
\end{equation}
(Here we have set $\beta_0=1$ for convenience.)  The weight 
\begin{align}
  w&=c_+^2+c_-^2=2\mathcal{N}^{-2}\notag\\
  &= \frac{\kappa^2-\gamma_s^4-\gamma_0^4-\gamma_0^2(1-2\gamma_s^2)}
  {(\gamma_s^2-\gamma_0^2)(1-\gamma_s^2+\gamma_0^2)},
\end{align}
reflects the dependence of the threshold of the breather on $\gamma_s$ and $\gamma_0$.   The larger $w$ is, the lower the threshold.  For example, Figs.~\ref{figs2}(b)--(c) respectively show $w$ versus $\gamma_0\in(0,\gamma_0^c)$ for $\gamma_s=\sqrt{7}/2$, and $w$ versus $\gamma_s\in(\gamma_s^c,\kappa)$ for $\gamma_0=0$. It shows that the threshold is increasing if $(\gamma_0^c-\gamma_0)$ or $(\gamma_s-\gamma_s^c)$ and thus $w$ become smaller.

\section{Comparison to a nonlinear Hermitian SSH model}

In the Hermitian nonlinear SSH model studied by Hadad \textit{et al.}~\cite{170622-1}, a self-induced topological end soliton can emerge when the lattice is driven at an end site.  However, the behavior of the soliton is different from the familiar topological end states of the linear SSH model.  Its energy is not pinned to zero, and rather than decaying exponentially away from the lattice boundary, it forms a ``plateau'' of almost constant magnitude.  The plateau feature seems to arise from an inability to form a topological domain wall between the boundary region and the lattice bulk.  Roughly speaking, if a domain wall were to form, it would induce a region of high intensity (corresponding to the topological end state), thus expanding the boundary domain and driving the domain wall deeper into the bulk.  By contrast, the end breathers in the present work are strongly localized, and clearly associated with self-induced topological domain walls.

The model studied by Hadad \textit{et al.}~differed from ours in one other respect: they used a Kerr nonlinearity rather than a saturable nonlinearity.  In this section, we show that the choice of nonlinearity is not responsible for the differences in behavior.

\begin{figure}
  \includegraphics[width=\linewidth]{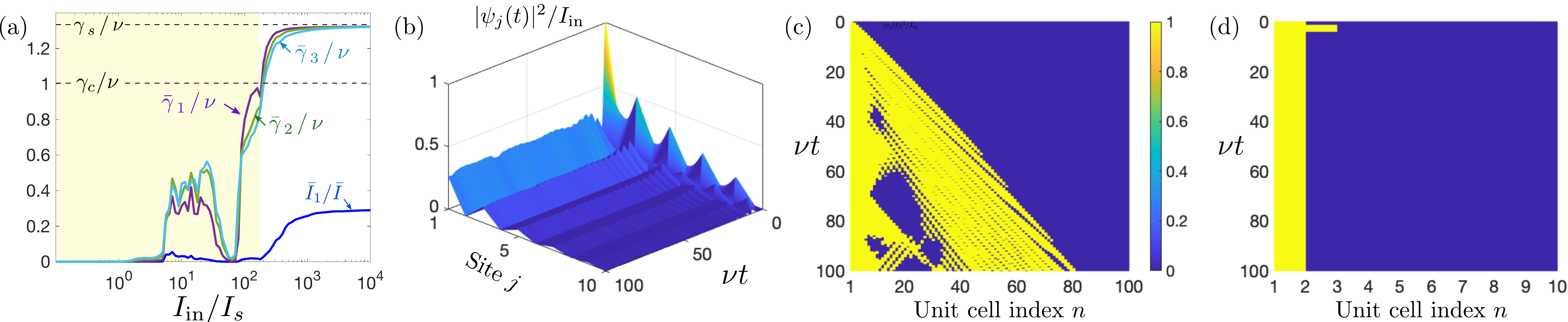}
  \caption{(a) Self-induced topological phase transition of the nonlinear SSH chain described in Eq.~\eqref{rec_Ham}, similar to Figs.~2(a) and 2(b) in the main text.  (b) Time-domain simulation results with the initial state $\Psi(0)=(\sqrt{I_{\rm in}},0,0,\cdots)^T$, using the parameters $\kappa = 2\nu$, $\gamma_0 = 0$, and $\gamma_s = (\sqrt{7}/2)\nu$.  Note that $\gamma_0$ and $\gamma_s$ are on the opposite sides of the critical point $\gamma_c = \nu$.  All other parameters are the same as in the Fig. 1(d).  (c) Heat map of the ``local'' topological phase, as characterized by $\Theta(\gamma_n-\gamma_c)$ (blue and yellow colors), versus unit cell index $n$ and elapsed time.  (d) Heat map of the ``local'' topological phase for the model described in the main text, using the parameters of Fig.~1(d) in the main text.}
\label{figs3}
\end{figure}

We consider the Hermitian Hamiltonian
\begin{align}
 H_\text{r}
   = \sum_{n = 1, 2, \dots} &\big[(\kappa-\gamma_n)a^\dag_{2n-1}  a_{2n} + \nu a^\dag_{2n}a_{2n+1}+\text{h.c.}\big],
\label{rec_Ham}
\end{align}
where the $\gamma_n$'s entering into the intra-cell hoppings have the saturable form given in Eq.~(2)--(3) of the main text.  This model is the same as that of Ref.~\cite{170622-1} except for the intensity dependence of the nonlinearity.  Unlike the model studied in the main text, the hoppings are all reciprocal (hence, there is no non-Hermitian skin effect).

Fig.~\ref{figs3}(a) shows the emergence of the self-induced end soliton using the Hamiltonian \eqref{rec_Ham}.  The detailed time domain simulation results, shown in Fig.~\ref{figs3}(b), reveal that the end soliton forms a plateau-like tail, consistent with the findings of Ref.~\cite{170622-1}.

Moreover, we can check the stability of the self-induced domain wall by examining the local values of $\gamma_n$ versus the critical value $\gamma_c = \kappa-\nu$.  As shown in Fig.~\ref{figs2}(c), the self-induced domain wall appears to have no unstable position; as time passes, it continuously extends deeper into the lattice.  Repeating this analysis for the model described in the main text, we find a domain wall that is stable in time, as shown in Fig.~\ref{figs3}(d).

\section{Mapping to a Creutz model with saturable gain/loss}
\label{a-map}

The SSH-like model with nonreciprocal and nonlinear intra-cell couplings, described in the main text, can be mapped via a unitary transformation to a Creutz model \cite{131104-2,171012-1}, shown in Fig.~\ref{figs1}.  The most important feature of the Creutz model is that the nonlinearities appear as on-site non-Hermitian terms---i.e., saturable gain and loss---rather than in the nonreciprocal hoppings.  This may be useful for experimental realizations of the phenomena described in the main text.

\begin{figure}[h]
  \includegraphics[width=0.3\linewidth]{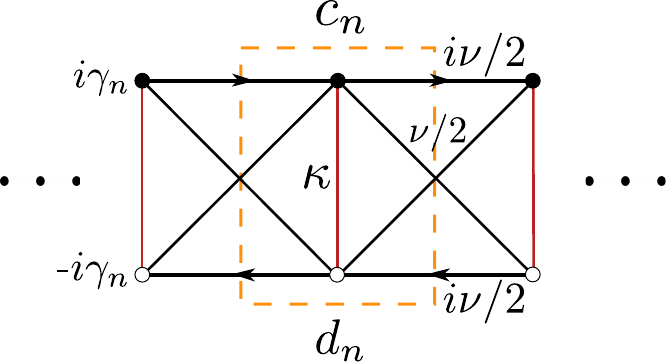}
  \caption{Schematic of a Creutz-like model with saturable on-site gain/loss.  The orange dashes indicate the $n$th unit cell, and $\pm i\gamma_n$ indicates the on-site gain/loss on chain $c/d$. The parameters are defined in the text.}
\label{figs1}
\end{figure}

To perform the mapping, we subject the nonlinear and nonreciprocal Hamiltonian $H$ given in Eq.~(1) of the main text to the basis change
\begin{equation}
  \begin{pmatrix}c_{n}\\d_{n}
  \end{pmatrix} = U \begin{pmatrix}a_{2n-1} \\a_{2n}
  \end{pmatrix}, \quad
  \begin{pmatrix}c^{\dag}_{n} & d^{\dag}_{n}
  \end{pmatrix}
  = \begin{pmatrix} a_{2n-1}^{\dag} & a_{2n}^{\dag}
  \end{pmatrix}
  U^{\dag} 
\end{equation}
in each unit cell $n$, where the unitary matrix $U = \exp(-i\sigma_x\pi/4)$ represents a $\pi/2$ rotation around the $x$ axis of the pseudo-spin-1/2 space ($\sigma_x$ denotes a Pauli matrix).  The transformed Hamiltonian is
\begin{equation}
  H'
  = \sum_n \Bigg\{\left[\frac{\nu}{2} \left(ic^\dag_{n}c_{n-1}
    -id^\dag_{n}  d_{n-1} +c^\dag_{n}d_{n-1}
    +d^\dag_{n}c_{n-1} \right)
    +\kappa c^\dag_{n}d_{n} + \mathrm{h.c.}\right]
  + i\gamma_n \left(c^\dag_{n}  c_{n}
  -d^\dag_{n}d_{n} \right) \Bigg\}.
  \label{nln_CHam}
\end{equation}
As shown in Fig.~\ref{figs1}, the Creutz model can be interpreted as two parallel 1D lattices with nonreciprocal nearest-neighbor intra-chain hoppings ($\pm i\nu/2$), nearest-neighbor inter-chain hoppings ($\kappa$) and next-nearest-neighbor inter-chain hoppings ($\nu/2$).  Moreover, the nonlinear $\gamma_n$ parameters appear as on-site gain and loss, on the $c$ and $d$ chains respectively.  In the linear regime $\gamma_n=\gamma$, this reduces to a non-Hermitian linear model with on-site gain and loss \cite{171012-1}.

We should also be careful to re-express the intensity $I_n$, on which $\gamma_n$ depends, in terms of the new variables.  This is achieved by taking
\begin{equation}
  \psi_{2n-1}(t) = \frac{1}{\sqrt{2}} [\varphi^{c}_n(t) + i\varphi^{d}_n(t)], \quad
  \psi_{2n}(t) = \frac{1}{\sqrt{2}} [i\varphi^c_n(t)+\varphi^d_n(t)]
\end{equation}
where $\varphi_n^{c/d}$ denotes the wavefunction on the $c$ or $d$ chain. We hence obtain the reasonable result
\begin{equation}
  \left|\psi_{2n-1}(t)\right|^2 + \left|\psi_{2n}(t)\right|^2
  = \left|\varphi^c_{n}(t)\right|^2 + \left|\varphi^d_{n}(t)\right|^2.
\end{equation}
The mapping to the Cruetz model also provides some insight into the stability of the topological end breathers.  The nonlinear parameter $\gamma_n$ is governed by the intensities on both the amplifying chain ($c$) and the lossy chain ($d$), in equal measure.

\clearpage
\end{widetext}

\end{document}